\documentclass[conference]{IEEEtran}
\usepackage{cite}
\usepackage{amsmath,amssymb,amsfonts}
\usepackage{algorithmic}
\usepackage{algorithm}
\usepackage{graphicx}
\usepackage{authblk}
\usepackage{textcomp}
\usepackage{xcolor}
\usepackage{braket}
\usepackage[hyphens]{url}
\def\BibTeX{{\rm B\kern-.05em{\sc i\kern-.025em b}\kern-.08em
    T\kern-.1667em\lower.7ex\hbox{E}\kern-.125emX}}

\begin{document}
\title{EQC : Ensembled Quantum Computing for Variational Quantum Algorithms}
\author[*]{Samuel Stein}
\author[$\dagger$]{Yufei Ding}
\author[*,$\ddagger$]{Nathan Wiebe}
\author[*]{Bo Peng}
\author[*]{Karol Kowalski}
\author[*]{Nathan Baker}
\author[*]{James Ang}
\author[*]{Ang Li}

\affil[*]{\small Pacific Northwest National Laboratory, Richland, WA, USA}
\affil[$\dagger$]{\small University of California Santa Barbara, Santa Barbara, CA, USA}
\affil[$\ddagger$]{\small University of Toronto, Ontario, ON, Canada}
 
\maketitle
\thispagestyle{plain}
\pagestyle{plain}

%%%%%% -- PAPER CONTENT STARTS-- %%%%%%%%

\begin{abstract}

Variational quantum algorithm (VQA), which is comprised of a classical optimizer and a parameterized quantum circuit, emerges as one of the most promising approaches for harvesting the power of quantum computers in the noisy intermediate scale quantum (NISQ) era. However, the deployment of VQAs on contemporary NISQ devices often faces considerable system and time-dependant noise and prohibitively slow training speeds. On the other hand, the expensive supporting resources and infrastructure make quantum computers extremely keen on high utilization.

In this paper, we propose a virtualized way of building up a quantum backend for variational quantum algorithms: rather than relying on a single physical device which tends to introduce temporal-dependant device-specific noise with worsening performance as time-since-calibration grows, we propose to constitute a quantum ensemble, which dynamically distributes quantum tasks asynchronously across a set of physical devices, and adjusting the ensemble configuration with respect to machine status. In addition to reduced machine-dependant noise, the ensemble can provide significant speedups for VQA training. With this idea, we build a novel VQA training framework called EQC that comprises: (i) a system architecture for asynchronous parallel VQA cooperative training; (ii) an analytic model for assessing the quality of the returned VQA gradient over a particular device concerning its architecture, transpilation, and runtime conditions; (iii) a weighting mechanism to adjust the quantum ensemble's computational contribution according to the systems' current performance. Evaluations comprising 500K times' circuit evaluations across 10 IBMQ NISQ devices using a VQE and a QAOA applications demonstrate that EQC can attain error rates very close to the most performant device of the ensemble, while boosting the training speed by 10.5$\times$ on average (up to 86$\times$ and at least 5.2$\times$). We will release EQC on GitHub.

\end{abstract}

\section{Introduction}
Quantum computing (QC) is poised to offer substantial computational capabilities that classical computing could never feasibly reach in many critical domains, such as database search \cite{grover1996fast}, graph combinatorial optimization \cite{farhi2014quantum}, quantum chemistry \cite{cao2019quantum}, machine learning \cite{biamonte2017quantum}, etc. Recently, Google has showcased quantum supremacy on their 53 qubits \emph{Sycamore} \emph{Quantum Processing Units} (QPU), where a quantum sampling problem, which would take an estimate of 10,000 years (rectified to 2.5 days by IBM) running on the ORNL Summit supercomputer now can be finished in $\sim$200 seconds \cite{arute2019quantum}.  

\begin{figure}[!htb]
\minipage{0.7\columnwidth}
    \includegraphics[width=\linewidth]{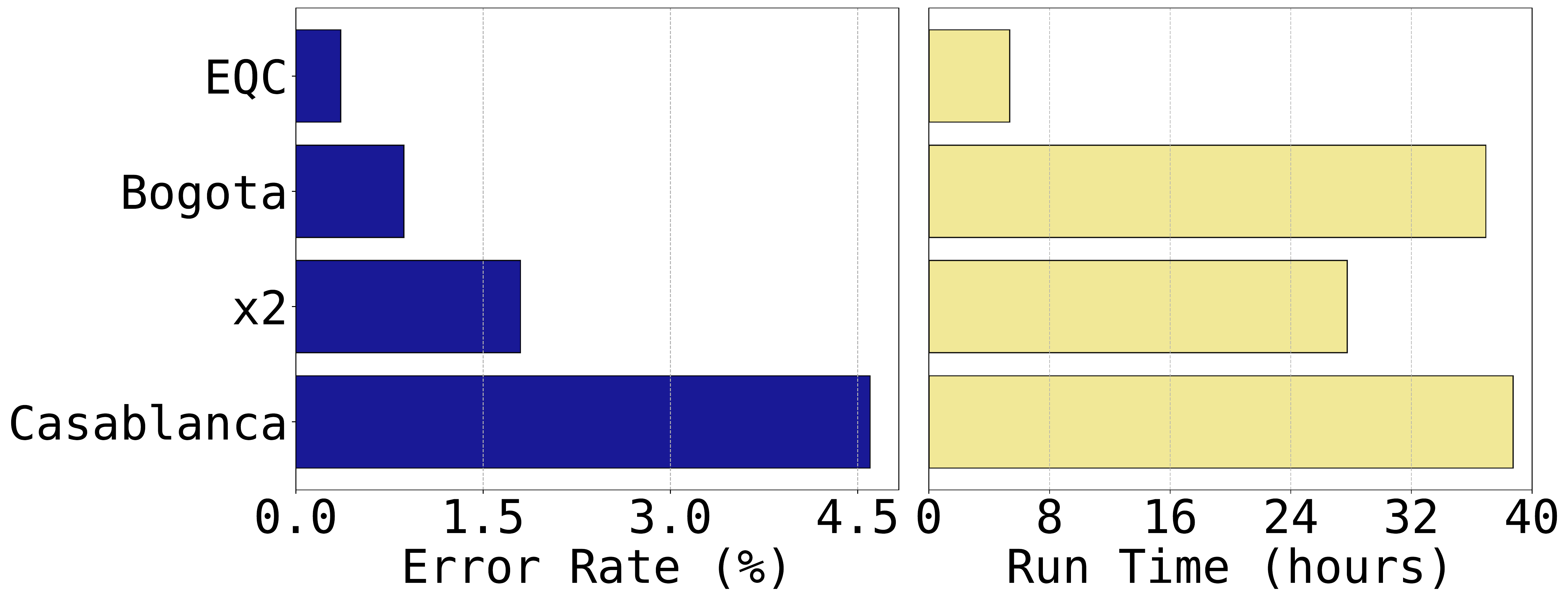}
\endminipage\hfill
\minipage{0.3\columnwidth}
    \includegraphics[width=2.1cm,height=2.5cm]{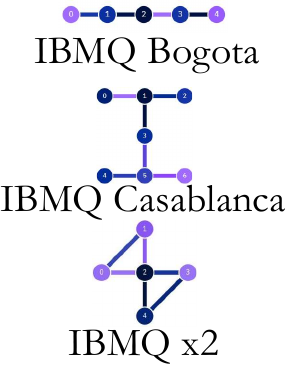}
\endminipage\hfill
\caption{VQE error rate, running time and QPU topology on IBMQ.}
\label{fig:run_time}
\end{figure} 

% \begin{figure}[!t]
%     \centering
%     \includegraphics[width=0.7\linewidth]{figures/machine_bias.pdf}
%     \caption{Machine Bias Error Rate}
%     \label{fig:error}
% \end{figure}

% \begin{figure}[!t]
%     \centering
%     \includegraphics[width=0.7\linewidth]{figures/running_time.pdf}
%     \caption{Running time}
%     \label{fig:run_time}
% \end{figure}
Although many algorithms that could provide quantum supremacy have been theorised \cite{grover1996fast, shor1999polynomial}, QPUs in the contemporary NISQ era are susceptible to noise in a multitude of ways. These include (i) \emph{Coherence Error}, which describes the natural decay of qubit states to their ground states, analogous to retention error in classical devices. Coherence error includes noise-induced state decay (i.e., T1 decay) and spin-spin relaxation (i.e., T2 decay); (ii) \emph{Gate Error}, which refers to imperfect gate operations. Such gate error is relatively small for 1-qubit gates (e.g., $\sim$0.1\% for IBMQ), but can be significant for 2-qubit gates (e.g., nearly 4\% for IBMQ \cite{tannu2019ensemble}); (iii) \emph{SPAM Error}, short for state preparation and measurement error, is introduced due to unsatisfactory state initialization and measurement sensing. Additionally, non-coherent interference such as cross-talk \cite{sheldon2016procedure, murali2020software} can also impose computational error, acting as device-specific latent noise. 

Due to the high error rate and insufficient qubits in supporting full-scale \emph{quantum error correction} (QEC), people are searching for quantum algorithms that can harvest the quantum advantages of NISQ devices but are more robust to the errors. The \emph{Variational Quantum Algorithm} \cite{cao2019quantum, cerezo2020variational, endo2021hybrid}, stands out as one of the most promising NISQ algorithms, which employs a classical optimizer to train a parameterized quantum circuit.

VQAs have been applied for several key applications. These include (i) the \emph{Variational Quantum Eigensolver} (VQE) for investigating molecular \& nuclear structures/dynamics in quantum chemistry \cite{yuan2019theory, arute2020hartree} and nuclear physics \cite{lu2019simulations, roggero2020quantum}; (ii) \emph{Quantum Approximate Optimization Algorithm} (QAOA) for optimizations such as MaxCut \cite{crooks2018performance} in graph analytics; and (iii) \emph{Quantum Neural Networks} (QNNs) for quantum machine learning \cite{peruzzo2013variational,farhi2014quantum,jiang2021co,stein2020qugan}.

Although VQAs have been extensively evaluated on platforms such as IBMQ \cite{castelvecchi2017ibm}, the deployment of VQAs for practical domain utilization (e.g., estimating the ground energy for molecules) on the NISQ platforms is still hampered by three challenges: (i) \emph{The training is susceptible to device-specific bias and noise}. Similar to the correlated error caused by mapping a circuit to the same qubits and links repeatedly following a fixed pattern \cite{tannu2019ensemble}, we have observed that QC is also subject to device-specific bias originating from topology, SPAM \cite{sun2018efficient}, calibration and running conditions. Figure~\ref{fig:run_time}-left shows the error rates of training a VQE circuit on three IBMQ devices (\texttt{Bogota}, \texttt{x2}, and \texttt{Casablanca}) individually with respect to ideal simulation, as well as the error reduction that can be achieved through our EQC framework; (ii)
\emph{Prohibitively long execution time.} Due to the significant deployment, maintenance, and operating costs, most QC platforms are provided as a cloud service and shared by many users. Given that VQA needs to repeatedly test \& adjust the parameterized circuit, iterate over enormous input data, train over numerous epochs, and wait for each trial going through the waiting queue, this leads to extremely long execution time. For example, it takes us about 4,623 hours to train a VQE circuit on Manhattan. The congestion, calibration and maintenance can further exacerbate the delay. Figure~\ref{fig:run_time}-middle shows the running time for our VQE circuit on the three IBMQ devices, compared to our ensemble approach; (iii) \emph{Large utilization variance due to unbalanced workloads.} Despite the sharing nature, quantum computers can be underutilized. This is largely due to workload imbalance across multiple devices given user-preference and congestion. As users tend to select the best performing quantum computer, the overall provider processor utilisation can be quite unbalanced. The VQA jobs significantly amplify such imbalance due to iterative testing and long execution time. Considering the expensive cost from the supporting infrastructure and sources such as cryogenic coolers, superconducting wires, microwave devices, and sophisticated control circuitry \cite{das2019case}, the under-utilization can be a considerable burden to the QC service providers. State of the art technologies such as Qiskit Runtime improve VQA performance through minimizing classical-quantum communication costs. However, these vertical approaches do not help with machine-specific bias, and may further exacerbate the unbalanced utilization problem.

\begin{figure*}[h]
    \centering
    \includegraphics[width=0.8\textwidth]{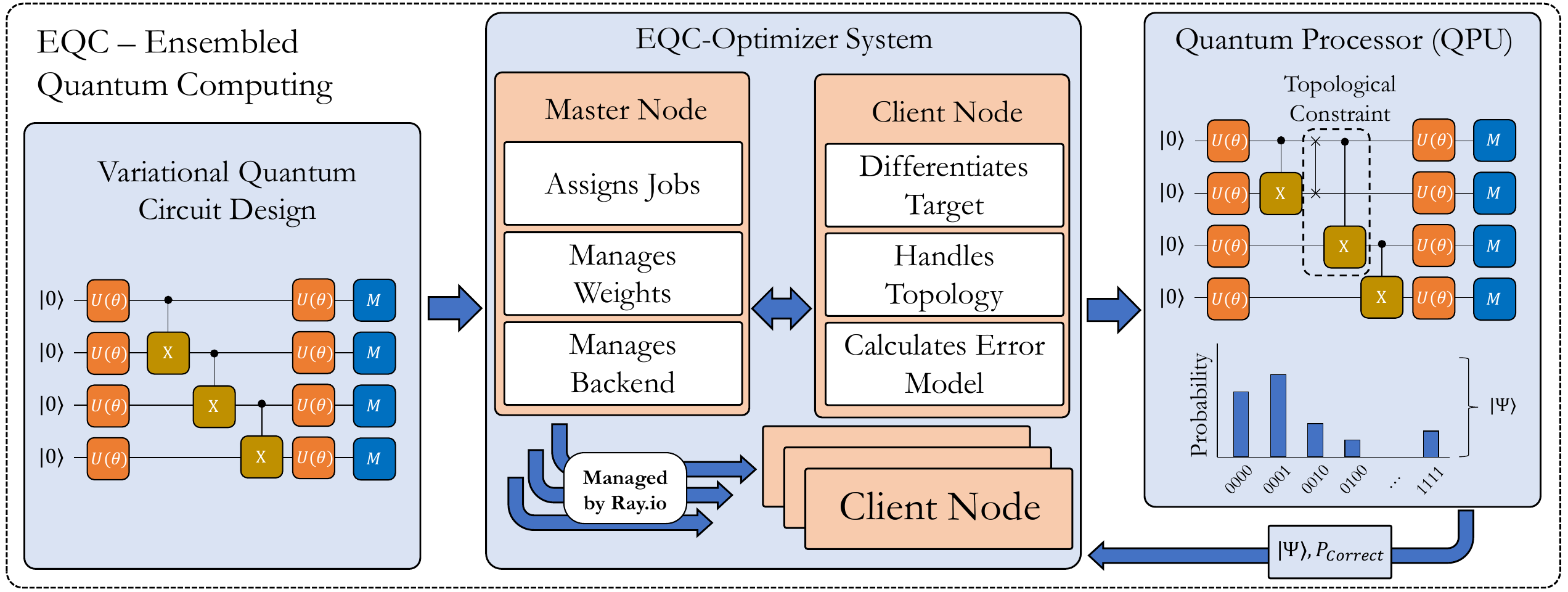}
    \caption{Overview of EQC framework. }
    \label{fig:section_guide}
\end{figure*}

In this paper, we propose EQC -- an Ensembled Quantum Computing framework for Variational Quantum Algorithms. EQC can bring considerable stability, noise-reductions and speed-ups in VQA training over alternative approaches. Being the primary effort in system-managed QPU-aware parallel VQA training, EQC employs a master node to asynchronously manage a set of client nodes paired with backend devices comprising the quantum ensemble. To dampen the QPU bound noise and minimize time-dependent machine drift, EQC further adopts an analytic model to assess the quality of the returned gradient, and builds a management module to allow online adjustment of the quantum ensemble based on the runtime condition of the backend devices. Evaluations on 10 IBMQ devices show that EQC can mitigate system-dependant bias while boosting the speed by over 10$\times$ (up to 86$\times$). This paper thus makes the following contributions:
\begin{itemize}
\item \emph{\textbf{The Concept and Design of Quantum Ensemble:}} we propose a novel way of thinking about a quantum backend: rather than using a single physical device incorporating device-specific bias and slow execution time, 
a quantum ensemble can serve as a virtualized backend for offering adaptive bias and noise mitigation through calibration-aware device mixtures, improved device choice relating to circuit design, and faster execution through parallelization.
\item \emph{\textbf{The EQC Framework:}} we develop a system architecture for asynchronous VQA optimization that can bring over an order of speedups over single QPU-based optimization, whilst simultaneously reducing error. We provide theoretic proof on the convergence of the approach.
\item \emph{\textbf{The Adaptive Weighting System:}} we propose a weighting approach considering noise level, topological constraint, transpiled circuit structure, calibration time, etc. It thus allows adaptive and dynamic weighting of the gradient with respect to the device condition, mitigating time-dependent drift and overwhelming time-dependant machine bias.
\end{itemize}

\section{Background}

\subsection{Quantum Gates and Devices}
QC exploits the quantum phenomena such as superposition and entanglement for computing, and manipulates the probabilistic state-vector to accomplish tasks. State-vectors, which represent the quantum states, are linear combinations of orthonormal eigenvectors with normalized square sum coefficients. Quantum computers make use of gates to perform state transformations, represented as unitary matrices relative to some basis, which operate on the state vector. Gate operations are performed in a physical quantum processing unit (i.e., QPU). There are two main features for QPUs: 

\vspace{2pt}\noindent\textbf{(i) Basis gates.} Basis gates describe the native hardware operations; all quantum circuits need to be transpiled to basis gates eventually in order to be executed by a QPU. For example, most IBMQ devices use the same basis gate set: \texttt{CNOT}, \texttt{ID}, \texttt{RZ}, \texttt{SX}, \texttt{X} \cite{cross2017open}. 
\begin{figure}
    \centering
    \includegraphics[width=0.4\textwidth]{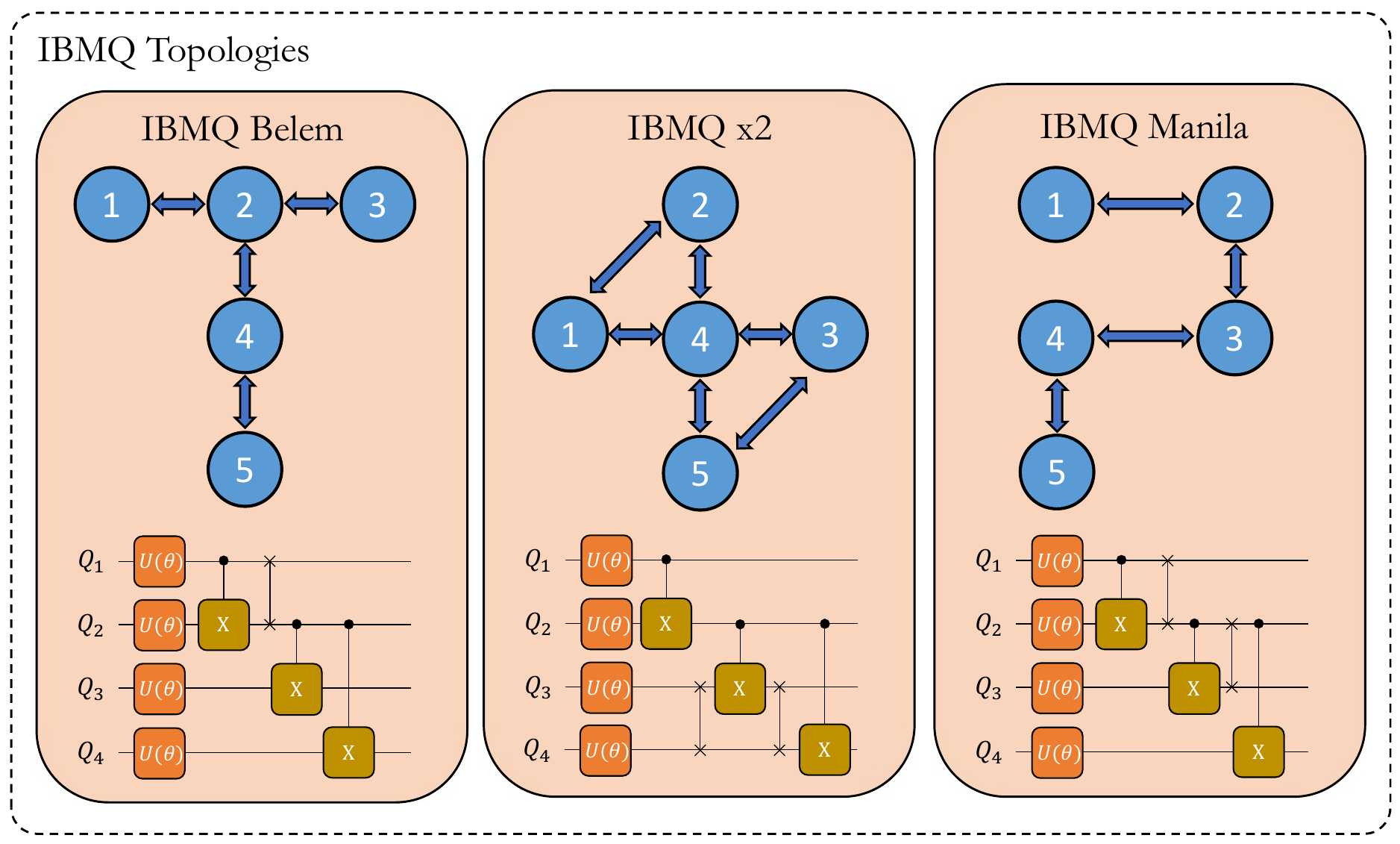}
    \caption{Topology illustration of 3 IBM Quantum Machines. Bi-directional connections indicate physical connections between qubits. Circuits illustrated perform the same computation transpiled to the respective topological constraints.}
    \label{fig:topological_graphic}
\end{figure}

\vspace{2pt}\noindent\textbf{(ii) Topology.} Applying two-qubit gates such as \texttt{CNOT} in theoretical development usually assumes fully connected systems. However, on real superconducting quantum processors, due to limitations on the physical designs, inter-qubit connectivity constraints exist. 

As shown in Figure~\ref{fig:topological_graphic},
the three devices (\texttt{IBMQ Belem},\texttt{x2} and \texttt{Manila}) all have 5 physical qubits. However, due to the variance of inter-qubit connectivity, the circuit structures for the same function can be different after transpilation. This is because the two-qubit gates such as \texttt{CNOT}, can only be performed on connected qubits; otherwise, the qubits must be moved next to each other using SWAP-gates. The swap gate is composed of three CNOT gates, a costly operation. The SWAP-gates allow any two qubits that are connected to swap positions, thus can hide the topology constraint, however at a substantial time and compounding error cost.

\subsection{QPU Noise}

\textbf{QPU Noise:} NISQ QPUs are highly diverse in their ability to run quantum routines without compounding a multitude of random errors due to low-level qubit stability and high-level QPU topological constraints. As discussed, there are three types of errors: (i) \emph{Coherence Error} due to the non-ideal energy exchange between the outside system and the ideal system, which is characterized by exponential decay factors $T1$ and $T2$, and known as thermal dephasing and decoherance \cite{georgopoulos2021modelling}; (ii) \emph{Gate Error} due to depolarization, which is characterized by undesired Pauli-X/Y/Z operations representing bit-flips or phase-flips.  (iii) \emph{SPAM Error} due to state preparation and measurement noise, which is associated with the anharmonicity between the states $|0\rangle$ and $|1\rangle$, and the erroneous process of distinctively and properly distinguishing between the two states correctly.

Additionally, each QPU has its own unique noise profile that changes with frequent calibration. These volatile systems vary in spatial and temporal noise due to the imperfect manufacturing process \cite{murali2020software}, imperfections of gate implementation and control, and specific external interference \cite{murali2019full, orrell2020sensor}. These noise compound with each other, increasing the overall probability of obtaining an erroneous outcome. 

\subsection{Variational Quantum Algorithm}

VQAs are characterized by a group of parameterized quantum operations, with parameters $[\vec{\theta} ]$, $\theta \subseteq \!R $. The quantum algorithms goal is to optimize  $[\vec{\theta}]$ to $[\vec{\theta^*}]$, where $[\vec{\theta^*}]$ is the optimal learned parameters, such that the trial wave function of $|\psi\rangle$ is transformed closer to some target wave function according to an optimization function. The optimization function represents for example the ground state of a quantum system, and is expressed as parameterized unitaries $U(\theta_i)U(\theta_{i-1})...U(\theta_1)|\psi\rangle$. For \emph{Variational Quantum Eigensolvers} (VQE) \cite{peruzzo2013variational} or \emph{Quantum Approximate Optimization Algorithm} (QAOA) \cite{farhi2014quantum}, the objective is to optimize towards some Hamiltonian $H$ describing the total energy of the system, or representing some objective cost function, as shown in Equation~\ref{eqn:equation_1}:
\begin{equation}
    argmin([\vec{\theta}])\,\langle\psi([\vec{\theta}])|\,\text{H}\,|\psi([\vec{\theta}])\rangle
    \label{eqn:equation_1}
\end{equation}
This optimization process is tailored for alternative use cases typically in quantum machine learning \cite{jiang2021co,schuld2014quest, stein2020qugan}.

The present VQA optimization process, however, encounters two key issues: (i) There is a substantial computational cost in inducing circuits and optimizing each individual parameter; (ii) Machine noise leads to a bias being learnt by a machine during training, where the parameters will deviate compensating for a machines noise. However, when a machine is re-calibrated or the backend device is changed, the learned parameters are no longer appropriate. Furthermore, if the noise is overwhelmingly high on a machine, the gradient computation will be erroneous and the algorithm will not learn.

\subsection{Distributed Optimization}
Distributed optimization of classical deep neural networks (DNNs) has been motivated by the need to train large neural networks \cite{brown2020language,krizhevsky2012imagenet} for exceedingly large datasets \cite{mannava2014overview} through parallel processing.
Asynchronous stochastic gradient descent (ASGD) has been shown to provide substantial speedups for DNN training with little accuracy loss \cite{zhang2013asynchronous}. However, the problem of distributing optimization across multiple classical processing units changes substantially when looking at a quantum distributed system. Classical systems primarily deal with latency, whereas quantum processors deal with both latency and individual noise characterization of each QPUs \cite{hamilton2020scalable}. The problem is exacerbated when multiple backend device architectures and/or providers are presented. In addition to the different runtimes of the service, the distinct periods of calibration, the diverse range of speed, as well as the noise can make the coordination of the training process significantly more complicated than classical distributive DNN training.

\section{EQC System Design}
In this section we describe the EQC architecture, structure  and discuss its implementation. 

%We refer the avid reader to works \cite{sweke2020stochastic,nedic2001distributed} for detailed proofs for our systems convergence.

\subsection{VQA Task Decomposition}
VQA optimization typically adopts stochastic gradient descent to perform step by step parameter differentiation, where concurrent tasks can be extracted from multiple hierarchies of the algorithm for (asynchronously) parallel execution.

Regarding \textbf{VQE}, these problems are commonly related to quantum chemistry, involving a Fermionic Hamiltonian that is decomposed via processes such as Jordan-Wigner decomposition. As a decomposed Hamiltonian is a linear sum of Pauli strings generated by transforming spin operators into a series of quantum computing supported Pauli operators. In this case, we perform the system parallelization at the Pauli string level. To compute the derivative of each respective parameter, the system collects parallelized computation, to perform one gradient descent step on one parameter. This is repeated for each parameter within the VQE ansatz.

Regarding \textbf{QAOA}, comprising of optimization problems such as MaxCut, the learning process aims to optimize over a target Hamiltonian objective. In the case of a single matrix representation of a Hamiltonian, the system aims to minimize or maximise $\langle \psi|H|\psi\rangle$. In doing so, we parallelize at the parameter level, where the derivative for each parameter is parallelized. This results in each parameter being asynchronously optimized, and is repeated over all parameters. Each parallel task is responsible for computing one gradient descent step over one parameter.

Regarding \textbf{QNN}, the learning process involves optimizing over $\frac{\delta f(x,\theta)}{\delta\theta} = \frac{1}{n}\sum_{i=1}^{n} \frac{\delta f(x_i,\theta)}{d\theta}$, where the average cost over a dataset is to be minimized. To parallelize the QNN training, we distribute at the dataset level for parameter optimization. Given a data set, as the parameter gradient is the average of all the data points' derivative for a parameter, the gradient can be parallelized for each parameter at each data point. Each parallel job is responsible for computing the gradient with respect to its assigned data point for a target parameter. One complete gradient descent step on one parameter is comprised of the average returned gradient for the target parameter over all data points, despite the gradients are applied asynchronously.

\subsection{EQC System Architecture}

As shown in Figure~\ref{fig:topological_graphic}, EQC is built as a single-master node, multi-client nodes architecture. It is designed for managing and optimizing the training of a VQA circuit with a set of parameters to optimize using a set of NISQ QPUs with different noise profiles, different topology's and different induced circuit architectures that may vary with time and transpilation. 
EQC consists of the master node, the client nodes, and the QPUs: the master node distributes workloads to client nodes, which perform target derivative calculation with their paired QPUs for their assigned task. With the computation of a derivative on a NISQ-term QPU comes a degree of confidence in said computation due to unavoidable noise. Therefore, the client node also computers a weight based on the transpiled circuit and reported system noise statistics at circuit induction time as well. The utilization of asynchronous stochastic gradient descent across a set of NISQ QPU's, whilst also incorporating a topology-aware low-level-performance-aware weighting system ensures the system attains a computational speedup, whilst improving system optimization performance.

\subsection{EQC Implementation}
We present the design details about the three major components of EQC: \emph{master node}, \emph{client node}, and \emph{QPU}.

\vspace{5pt}
\subsubsection{Master Node}
The master node is provided with three key pieces of information: (i) the quantum circuit over which the system is to be optimized with measurements; (ii) the initialized parameters $[\vec\theta]$, $\theta\subseteq R$; and (iii) a detailed loss function $\ell$ describing the optimization problem.

The master node is responsible for keeping track of all parameters at all times, as well as the ideal (i.e. no topological constraints) circuit layout. Once the master node has been initialized with all of the pre-requisites in place, the master node queries the Quantum Computing service provider(s) based on the ensemble configuration. The available QPUs to form the ensemble should have active qubits larger than the number of qubits required by the parameterized circuit. The backend devices can be homogeneous or heterogeneous. In other words, although we demonstrate EQC using 10 of IBMQ devices, the framework itself supports an ensemble of QPUs from different vendors. 

When the master node dictates the formalization of the ensemble, it initializes client nodes one per device with the loss function and circuit template. The master node then distributes the VQA task parameters in a cyclic fashion to the client nodes. The master node adopts a task queue for parallel task distribution. The master node iterates the parameter list and asynchronously assigns the next parameter to a client node that is available until no parameters left. After that, the next epoch can start.

\begin{algorithm}[!t]
\caption{EQC Master Node}
\label{alg:master}
\begin{algorithmic}
\STATE Set $\epsilon_{\text{END}}$,\text{ }$\alpha$ and $\ell$
\STATE $C \gets$ Circuit Template 
\STATE Parameters $\gets [\vec{\theta}]$ \\
\STATE $\epsilon,i \gets 0$
\STATE $K \gets [QPU's]$
\STATE $G \gets [\text{k Client Nodes}]$
\FOR{QPU, Client in $K,G$} 
\STATE $Client  \gets  QPU,C,\text{ and } \ell([\vec{\theta}])$
\ENDFOR

\FOR{Param in Parameters}
    \STATE $G_i \gets \text{Compute } \frac{\delta\ell}{\delta\theta_i}$ \\ 
    \STATE $i \gets i+1$
    \IF{$i > len([\vec{\theta}])$}
    \STATE $i \gets 0$
    \ENDIF
\ENDFOR
\WHILE{$\epsilon < \epsilon_{END}$}
\FOR{CN in Ready Clients}
\STATE $\frac{\delta\ell}{\delta\theta_i},P_{\text{Correct}} \gets CN$
\STATE $P_{\text{Correct}} \gets Bound(P_{\text{Correct}})$
\STATE $\theta_i =\theta_i - P_{\text{Correct}}(\alpha\frac{\delta\ell}{\delta\theta_i})$
\STATE $CN \gets [\vec{\theta}]_{\text{NEW}}$
\STATE $CN \gets \text{Calculate } \frac{\delta\ell}{\delta\theta_i}$ 
\STATE $i \gets i+1$
\IF{$i > len([\vec{\theta}])$}
\STATE $i \gets 0$ \\ 
\STATE $ \epsilon \gets \epsilon + 1$
\ENDIF
\ENDFOR
\ENDWHILE
\end{algorithmic}
\end{algorithm}

The only information the master node receives from the client nodes are the gradients of the parameters $\theta_i$, and the computed noise model of the QPU associated, which we will discuss in the following section. On receipt of a gradient, the master node applies this gradient using the ASGD rule outlined in Equation~\ref{eqn:asgd}, and sends a new parameter to differentiate at an idle client. The whole process is outlined in Algorithm~\ref{alg:master}.

\vspace{5pt}
\subsubsection{Client Node} 
The primary objective for the client nodes is to maximize the usage time of each QPU available. Each client node is responsible for managing one QPU's experiments and maximizing its utilization. On initialization, each client node is provided with the quantum circuit template, the unique id, and the optimization function $\ell$. 

The client node then receives a parameter index to differentiate and generates the forward and backward pass from the parameter shift rule.
The client node has the information about the topological constraints of the resident QPU, as well as all QPU noise
factors such as T1, T2 and CNOT error rates. Using the topological constraints, the client node transpiles the template circuit with respect to the target QPU topology. The client node runs the transpiled forward and backward circuit in the QPU, and processes the resultant two probability distributions through the cost function, attaining $\frac{\delta\ell}{\delta\theta_i}$.

Finally, the client node returns the obtained gradient and an associated noise factor back to the master node, and in return acquires the next pending job to repeat this process. The procedure of the client node is shown in Algorithm~\ref{alg:client}.

\vspace{5pt}
\subsubsection{QPUs} Each quantum processor has its own topological constraints, architecture type and noise factors (see Table~\ref{tab:testbed}). The topology can be represented by a non-directed graph, as shown in Figure~\ref{fig:topological_graphic}. To accommodate such topological constraints, extra SWAP operations are needed. With respect to noise, each QPU is substantially different. Highly connected QPU's suffer from cross-talk, degraded CNOT-gate fidelity and T1/T2 lifespans. As a comparison, newer architectures such as the IBM Falcon processor are laid out in a honeycomb fashion to mitigate the cross-talk issue and reduce noise at the cost of a lower connectivity. Nevertheless, noise still applies to each NISQ system, and is correlated to the topological constraints considering SWAPs and circuit depth.

\begin{algorithm}[!t]
\caption{EQC Client Node}
\label{alg:client}
\begin{algorithmic}
\STATE $C \gets$ Circuit Template
\STATE $Q \gets$ Connect to QPU
\STATE $C_{\text{Transpiled}} \gets Transpile(C,Q)$

\STATE $[\vec{\theta}] \gets$ Parameters
\WHILE{Calculate $\frac{\delta\ell}{\delta\theta_i}$}
\STATE $[\vec{\theta}]$ $\gets$ $[\vec{\theta}]_{\text{NEW}}$
\STATE $[\vec{\theta}]_{\text{FWD,BCK}} \gets [\vec{\theta}],[\vec{\theta}]$
\STATE $[\vec{\theta}]_{\text{FWD,BCK}} \gets [\vec{\theta_i}] \pm \frac{\pi}{2}$
\STATE  Job $\gets$ Submit $C_{Transpiled}([\vec{\theta}])_{\text{FWD,BCK}}$
\STATE $P \gets P_{\text{Correct}}(C_{\text{Transpiled}},QPU)$
\IF {Job is Ready}
\STATE $|\Psi\rangle_{\text{FWD,BCK}} \gets \text{Job Results}$
\STATE $\frac{\delta\ell}{\delta\theta_i} = \frac{\ell(|\Psi\rangle_{\text{FWD}} - \ell(|\Psi\rangle_{\text{BCK}}}{2}$
\STATE Ready($\frac{\delta\ell}{\delta\theta_i},P_{\text{Correct}}$)
\ENDIF
\ENDWHILE
\end{algorithmic}
\end{algorithm}

%With regards to IBM-Q, the major platform for this work, circuit induction is comprised of queuing time and circuit induction time. Quantum devices on IBMQ can require between a second to a day of queuing time to induce a circuit. IBM-Q operates on a token-based authentication mechanism integrated into the Qiskit Python API, which also allows for collecting the information of machine statistics, queuing time, topological constraints and calibration. These allow the users to manually manage the distribution of workloads according to device properties and runtime conditions.

\subsection{Implementation Details}
EQC is implemented based on two  packages: \emph{Qiskit} and \emph{Ray}. Qiskit is responsible for circuit generation, communication management with IBMQ devices, and all other quantum related processes. Meanwhile, Ray \cite{moritz2018ray} provides very flexible and universal APIs for building distributed systems. Ray was originally designed for distributive reinforcement learning. Ray.io creates remote actors generated based on Python's class structure, which are responsible for managing the QPU-related configuration information, and properties mentioned above with respect to client node responsibilities. Ray.io allows for asynchronous communication between the master node and client nodes, through the use of "ready" calls that indicate the client node has a pending object to be collected. Each client node connects to one running, unreserved QPU by connecting to individual IBMQ backend.

\begin{figure}
    \centering
    \includegraphics[width=0.4\textwidth]{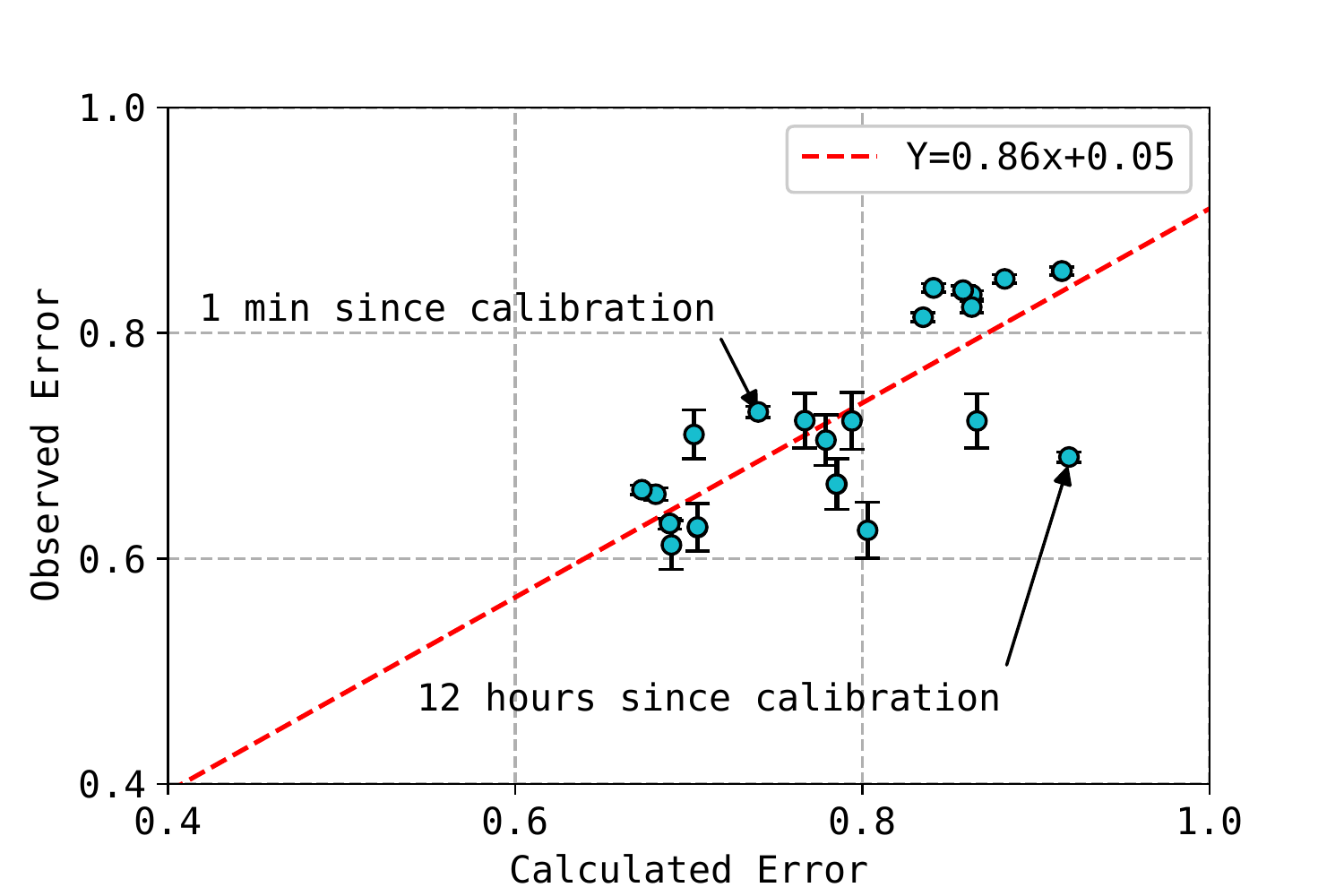}
    \caption{Calculated Vs Observed 5-Qubit GHZ State Error. Observed error indicates proportion of erroneous outputs, and calculated error indicates the computed chance of an error. }
    \label{fig:error_model}
\end{figure}

\section{Worker QPU Weighting}

As discussed, NISQ devices are extremely noisy and compound computational errors in a multitude of ways. Furthermore, their noise models change at every calibration. Certain quantum computers are better at producing more accurate resultant probability distributions than others, due to coherent and latent system noise factors. In order to take these noise factors into consideration for EQC, we propose to quantify and weight the gradient based on an analytic model.

%As discussed in \cite{georgopoulos2021modelling}, 

A key observation that aligns system noise (measurement, depolarization, and thermal/dephasing decay) here is that they can be attributed as probabilistic events. Each quantum computer, when calibrated, reports the gate fidelity, measurement fidelity, gate times, state anharmonicity, and T1/T2 decay constants. The event of the identity gate being applied (i.e. no error) is $1-p$ for each gate fidelity reported. For thermal decay, the exponential mode of $P(\Phi_{x,t} = \Phi_{0,t}) = e^{-t/T1}$ where $\Phi_{x,t}$ represents the probability of a decay occurring between time $0$ and $t$. The probability of a error-free computation occurring thus can be expressed as the product of possible errors not occurring. Therefore, we propose characterizing an entire systems quality by compounding these calibration statistics based probabilistic models with quantum routine properties described below:
\begin{equation}
    P_{Correct} = e^{\frac{-CD\frac{\mu_{t-G1}+\mu_{t-G2}}{2}}{T_1T_2}}(1-\gamma)^{G_1}(1-\beta)^{G_2}(1-\omega)^M
    \label{eqn:p_model}
\end{equation}
where CD refers to the critical depth of a circuit, $\mu_{t-G_{1/2}}$ is the average gate time of a 1 or 2 gate operator, $\beta$ is the CNOT fidelity, $\gamma$ is single gate fidelity, $G_{1/2}$ is the number of single or dual gate operations, $\omega$ is the measurement error rate and $M$ is the number of measurements. Since this equation is a probabilistic model, it is bound by $0\leq P_{Correct}\leq 1$. Furthermore, our model benefits from being topology-aware. By incorporating dual gate count $G_2$, topological constraints will drive this value up due to increased SWAP gates increasing probability of error, thereby decreasing weights of computationally consequential topologys.

To validate our analytic model, we apply a 5-qubit GHZ state onto real quantum computers and use our model to predict the probability of error. The GHZ state is characterized by the state vector $|\Psi\rangle = (\frac{1}{\sqrt{2}}(|00...0\rangle + |11....1\rangle)$. Therefore, any results with both a 0 and 1 in the bit string have had a computational error occur. We compute our expected $P_{Correct}$  and compare with IBMQ's QPU's \texttt{IBMQ Lima},  \texttt{IBMQ x2}, \texttt{IBMQ Belem}, \texttt{IBMQ Quito}, \texttt{IBMQ Manila} and \texttt{IBMQ Bogota} in Figure \ref{fig:error_model}. 

\begin{figure}
    \centering
    \includegraphics[width=\linewidth]{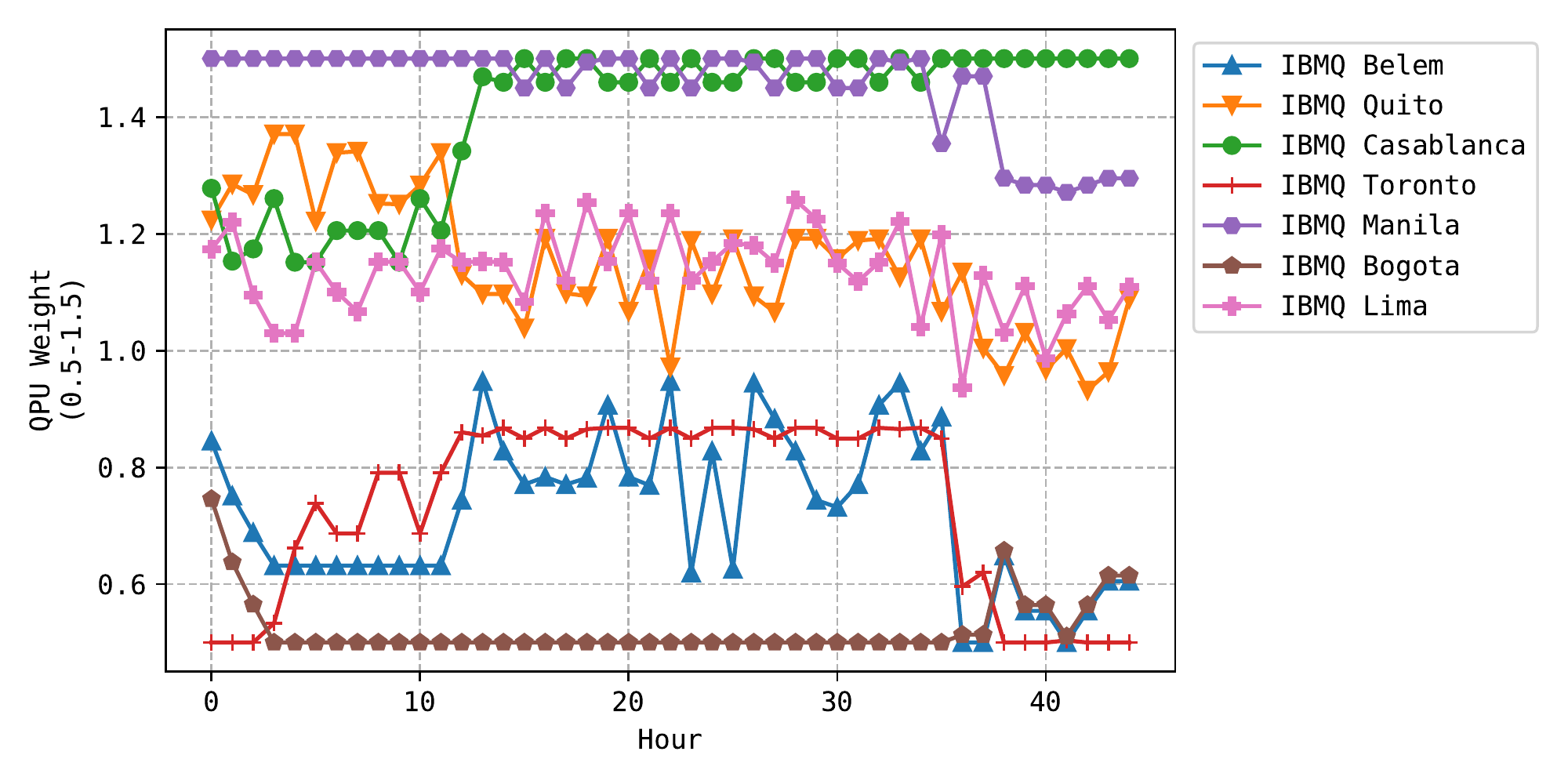}
    \caption{QPU Weighting: Weighting system applied with weights bound $[0.5,1.5]$ over 7 QPU's transpiling and computing Equation \ref{eqn:p_model} over circuit illustrated in \ref{fig:vqe_circuit}} 
    \label{fig:temporal_qpu_weighting}
\end{figure}

As can be seen in Figure \ref{fig:error_model}, a strong linear positive correlation is demonstrated between observed errors and $P_{Correct}$ with a $R^2$ value of $0.605$. Furthermore, we calculate the Pearson correlation between our calculated errors and observed errors. A correlation of 0.784 is attained with a two-tailed p-value of 1.28E-7, indicating that our results and actual error rates are strongly statistically correlated. We observe that our model was substantially better at predicting the chance of an error closer to calibration times. As observed in Figure \ref{fig:error_model}, for a freshly calibrated system we observed an error rate of $73.5\%$, alongside a calculated error rate of $74.0\%$. However, for staler calibrations we observe an error rate of $69.0\%$ alongside a calculated error rate of $91.9\%$. This was only observed on certain machines however, with other machines not suffering from the same problem of stale calibrations. We make use of our characterizing formula by giving each worker node the job of calculating its $P_{Correct}$ at transpilation before each circuit induction. We make use of our model by linearly scaling the distribution of $P_{Correct}$ values to be bound by a range. This weighting system is real-time adaptive, and incorporates the transpiled cost of a circuit and the QPU's stability. We demonstrate this in Figure~\ref{fig:temporal_qpu_weighting} where a group of IBMQ machines are bound to have weights between $[0.5,1.5]$. We demonstrate the ability to live adapt to noise changes, and variance in transpilation to backends that might result in inferior performance. 
\section{EQC Evaluation}
\begin{figure*}[!t]
    \centering
    \includegraphics[width=1\textwidth]{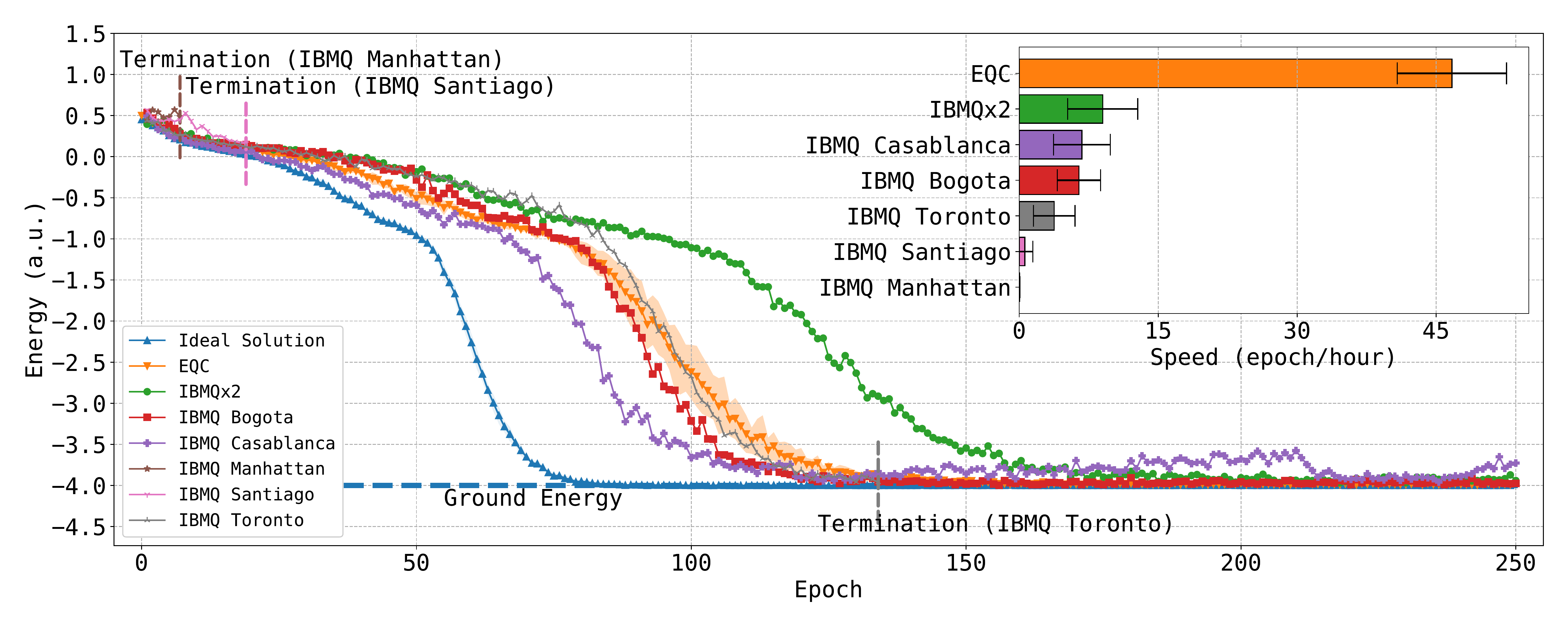}
    \caption{4-Qubit Heisenberg model on a square lattice. Termination represents an experiment being terminated beyond 2-weeks of running time. Error bars are present for EQC and the Ideal Solution. Independent machine training is represented by a machine name in the legend. }
    \label{fig:results}
\end{figure*}
\begin{table}[ht]
\footnotesize
\centering
\caption{IBMQ platforms used for evaluation. QV refers to quantum volume \cite{cross2019validating}. The line, T-shape and fully-connected topology can be seen in Figure~\ref{fig:topological_graphic}.}
\centering
\begin{tabular}{c|c|c|c|c}
\hline
\textbf{Device}    & \textbf{Qubits}               & \textbf{Processor}      & \textbf{QV}        & \textbf{Topology}     \\ \hline 
Lima       & 5                          & Falcon r4T    & 8  & T-shape                     \\ \hline
x2     & 5                          & Falcon r4T    & 8  & Fully-connected                     \\ \hline
Belem   & 5                          & Falcon r4T     & 16    & T-shape                \\ \hline
Quito   & 5                          & Falcon r4T      & 16   & T-shape                \\ \hline
Manila       & 5                          & Falcon r5.11L     & 32    & Line                 \\ \hline
Santiago      & 5                          & Falcon r4L      & 16    & Line                   \\ \hline
Bogota    & 5                          & Falcon r4L     & 32     & Line                \\ \hline
Lagos       & 7                          & Falcon r5.11H      & 32  & H-shape                     \\ \hline
Casablanca     & 7                          & Falcon r4H     & 32   & H-shape            \\ \hline
Toronto     & 27  & Falcon r4      & 32  & Honeycomb                   \\ \hline
Manhattan     & 65  & Falcon r4      & 32  & Honeycomb                   \\ \hline

\end{tabular}
\label{tab:testbed}
\end{table}

\begin{figure}
    \centering
    \includegraphics[width=1\linewidth]{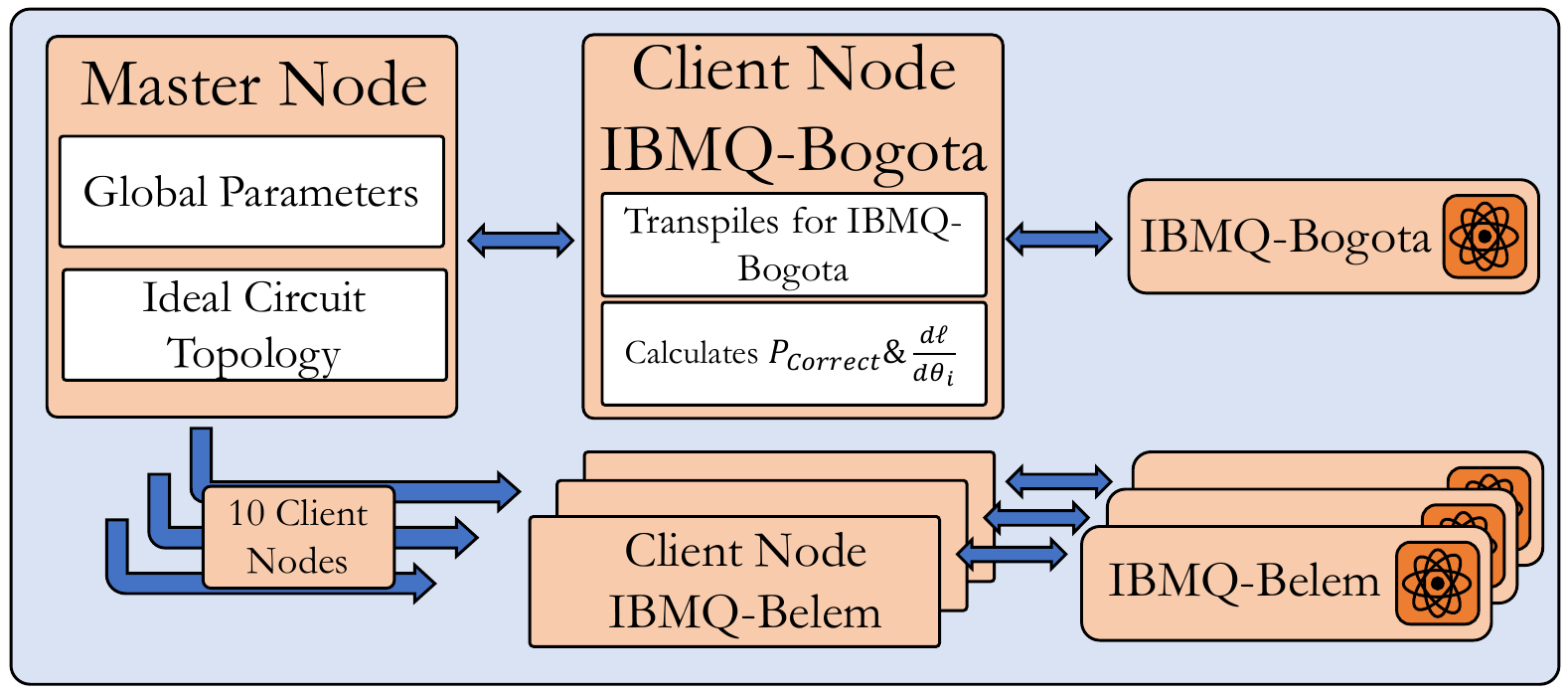}
    \caption{Experimental Setup}
    \label{fig:exp_set}
\end{figure}

\subsection{VQA Problem Settings}
 We comprehensively evaluate EQC using the Variational Quantum Eigensolver optimization problem, as well as the Quantum Approximate Optimization Algorithm. The experimental setup is illustrated in Figure~\ref{fig:exp_set}. We use up to 10 IBMQ machines in deploying and evaluating our system, listed in Table~\ref{tab:testbed}. In total, our evaluations for this work comprise $\sim$500,000 circuit executions on IBMQ devices.  
 
 \subsection{Variational Quantum Eigensolver Evaluation}
 The target problem is an application to quantum magnetism, whereby we perform the minimization of a 4-qubit Heisenberg model Hamiltonian on a square lattice \cite{kandala2017hardware}. The Hamiltonian is described by:
\begin{equation}
    H = J\sum_{(i,j)}(X_iX_j + Y_iY_j + Z_iZ_j) + B\sum_iZ_i
    \label{eqn:hamiltonian}
\end{equation}

\begin{figure}
    \centering
    \includegraphics[width=0.85\linewidth]{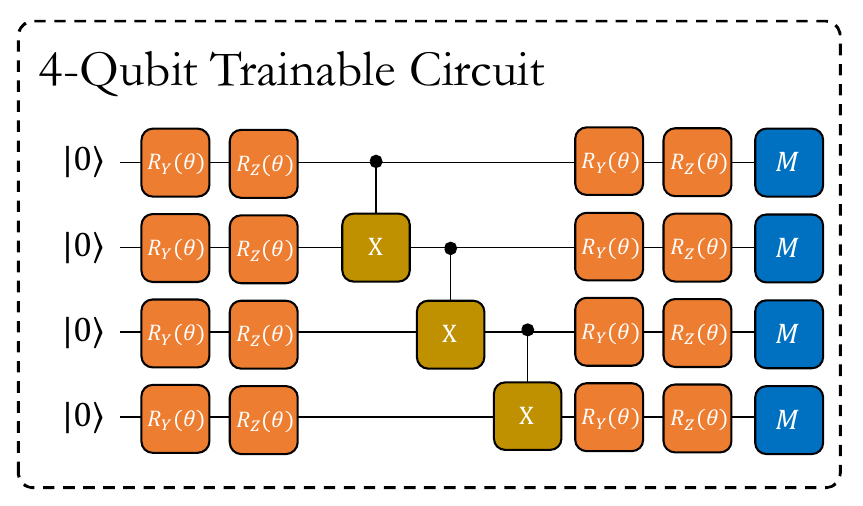}
    \caption{4-Qubit VQE Circuit. X gates connected to another qubit represent CNOT gates. M gates represent measurements.}
    \label{fig:vqe_circuit}
\end{figure}

\noindent where $J$ refers to the spin-spin strength and $B$ is the magnetic field along the Z-axis. We evaluate the Hamiltonian that has $J/B=J=B=1$, the same experiment as already demonstrated in \cite{kandala2017hardware}, which provides a reliable baseline (i.e., ground energy) for verifying the convergence. The 4-qubit square lattice results can be represented as a graph  with $V=[1,2,3,4]$ and edges $E=[(1,2),(2,3),(3,4),(1,4)]$. We make use of a hardware-efficient ansatz comprised of a linearly entangled full-bloch sphere rotation style circuit. The parameterized circuit is visualized in Figure~\ref{fig:vqe_circuit}, composed of $R_Y$, $R_Z$ and \emph{CNOT} gates. This circuit design exploits a full bloch-sphere rotation in the initial two gate layers of $R_Y$ and $R_Z$ gates, manipulating both phase and probability amplitudes. This is followed by each qubit being linearly entangled with one and other (i.e. CNOT(1,2), CNOT(2,3), ... (CNOT(n-1,n)), followed by another two layers of gates $R_Y$ and $R_Z$. This is similarly motivated to the circuit design used in \cite{kandala2017hardware}. We make use of parameterized differentiation according to the ASGD update rule at a learning rate ($\alpha$) of 0.1. To incorporate node weighting, we make use of Equation \ref{eqn:weighted_asgd}.

\begin{equation}
    \theta^{t+1}_i = \theta_i^t - P_{\text{Correct}}\alpha g^\tau (\theta^\tau_i)
    \label{eqn:weighted_asgd}
\end{equation}

\subsection{VQE Unweighted Performance Evaluation}

We show how EQC can contribute to noise-mitigation and machine-bias mitigation, while gaining significant performance improvement over single-machine optimization. The primary results are shown in Figure~\ref{fig:results}. We show the optimization process of minimizing the system energy as described in Equation~\ref{eqn:hamiltonian} for over 250 epochs. The ideal learnt ground state is illustrated at $-4.0$ a.u. (as validated in \cite{kandala2017hardware}). We further illustrate the average single machine speed in comparison to EQC. We evaluate the optimization of our system on an ideal quantum simulator using 8192 shots with an average convergence point of 80 epochs over 10 experiments. This serves as the baseline as plotted in Figure~\ref{fig:results}. Notably, the standard deviation is plotted, however the value is negligible and hence is difficult to see on the plot. We further run the optimization problem once over 6 single IBMQ machines, each attempting to train the system on their own independent of each other. The optimization task is run once on single machines due to severe time costs of single-machine VQE training. 

Overall, \texttt{IBMQ x2}  is shown to be the slowest at converging with an approximate convergence at 175 epochs, a 118.75\% increase in the number of epochs to converge. The relatively slow convergence is attributed to \texttt{IBMQ x2}'s older topological architecture and high degree of crosstalk, therefore being substantially more error prone. In comparison, for \texttt{IBMQ Bogota}, we observe convergence substantially quicker at epoch 122, 52.5\% slower than an ideal simulator. Certain experiments were infeasible to complete, where in IBMQ Manhattans, running a 250-epoch 16-parameter VQE requires at the order of months to train, with our current expectation being 193 days. Similarly, Santiago would be on the order of weeks, with our approximation being 21 days. These numbers are also susceptible to large time-dependent fluctuations, with Toronto fluctuating from 6.553 epochs per hour to 0.033 epochs per hour over our entire training period.

Due to these large swings in performance, we terminate certain experiments which had not finished after 2 weeks of training, as seen in Figure \ref{fig:results}. \texttt{IBMQ Casablanca} was substantially more performant at an average performance of 6.767 epochs per hour, hence only a run time of 36.944 hours, and converging at epoch 130. Although Casablanca was most performant initially in the epochs between 0 and 130, we observe the machine becoming unstable until epoch 215 where it stabilizes back at the ground energy. This is an example of time dependent drift whereby the machine becomes noisier over time, and the learned parameters are biased, compensating for this noise, and avoiding the actual optimal solution.

As demonstrated, two key factors that are observed in a single machine training: machine throughput and machine stability, which are highly variable with time. 
\begin{figure}
    \centering
    \includegraphics[width=1\linewidth]{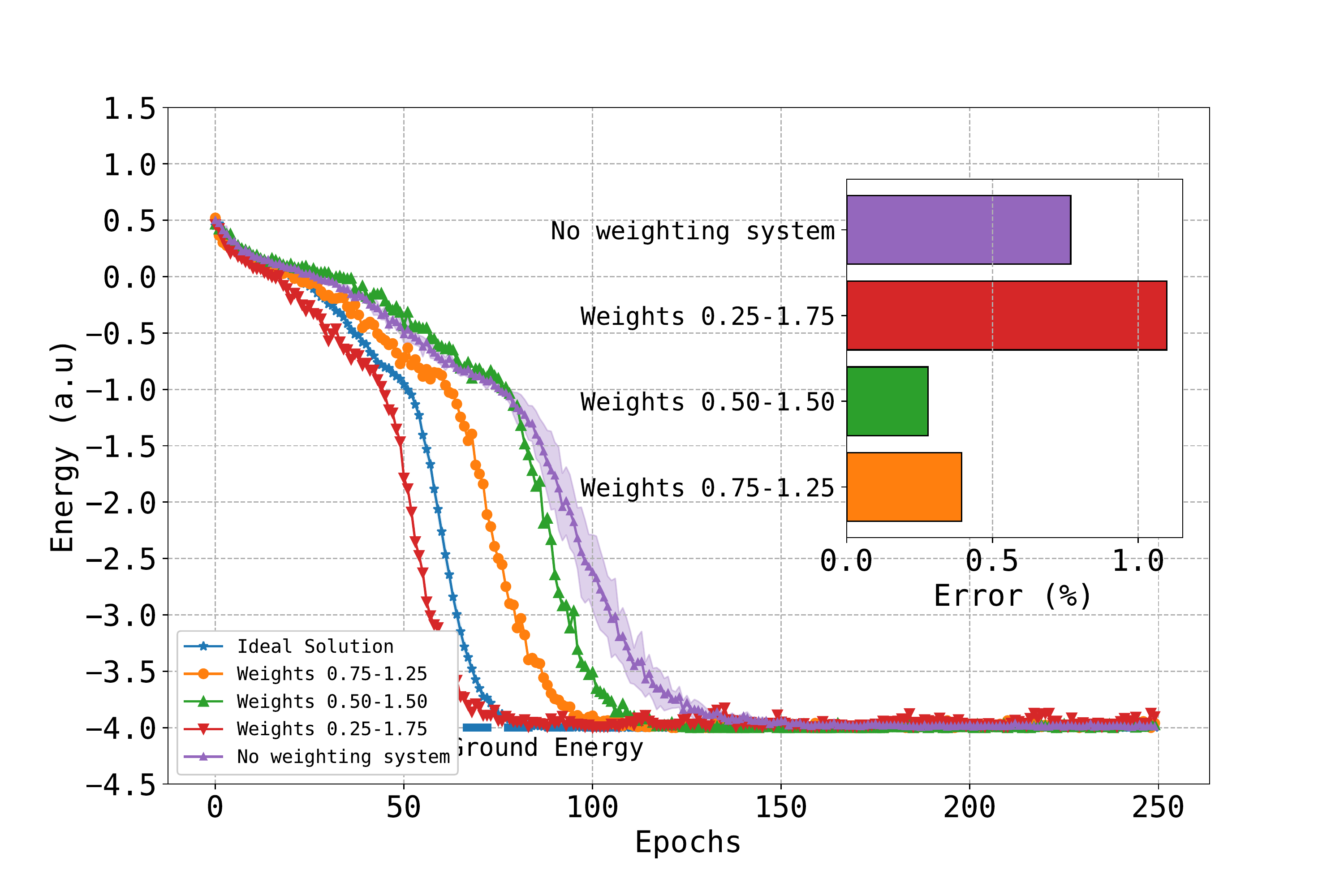}
    \caption{4-Qubit Heisenberg Weighted QPU Results. Results from EQC are evaluated using the update rule in eqn-\ref{eqn:weighted_asgd}, where no weights equates to $w=1$. Error ($\%$) represents average error rate relative to ground energy of -4.0 a.u.}
    \label{fig:results_weighted_vqe}
\end{figure}

We evaluate the same problem on EQC, as represented in Figure \ref{fig:results}. We observe an average speed of 46.701 epochs per hour when using our system over 10 IBMQ machines. The fastest IBMQ machine is \texttt{IBMQ x2}, performs at a rate of 9.011 epochs per hour. With EQC providing a worst-case improvement of 518\%. We train our system on EQC 3 times and illustrate the standard deviation at each epoch on Figure~\ref{fig:results}. EQC converges at epoch 135, 69\% slower than an ideal simulator. This is substantially better than the noisiest machine \texttt{IBMQ x2}, which also is the fastest machine and hence contributed the largest throughput to EQC.
One of our key insights that we demonstrate is the ensemble noise mitigation and machine-bias mitigation of EQC. With EQC being an ensemble of QPU's each with their own machine specific noise profile the problem of large machine-induced bias is dampened inherently through the mixture, i.e., an ensemble of averaged noise. 

Quantum processors tend to have time dependent drift from the ideal parameters. For example, 
Casablanca converges the fastest, however drifts from the ideal solution after convergence, as shown in Figure~\ref{fig:results}.This dampening is demonstrated by the final average converged energy of each system. With an ideal ground energy of $-4.0$, the error of a system is the average of the obtained ground energy divided by the ideal ground energy. The deviation from ground state error is visualized in Figure~\ref{fig:run_time}-right. As can be seen, EQC attains an error rate of $0.379\%$, whereas \texttt{IBMQ x2} attains an error rate of $1.798\%$, Bogota $0.865\%$, and Casablanca $4.6\%$. We can see that EQC attains the closest result to the ideal solution when compared with training on single machines, implying that the ensemble can mitigate device-specific noise. 

\subsection{VQE Client Node Weighting Evaluation}
We evaluate EQC against two different weighting strategies, and compare them to baseline (i.e., no weighting) and the ideal simulation result. As shown in Figure~\ref{fig:results_weighted_vqe}, the application of our weighting model can allow for an improvement in the rate of convergence. The weighting system operates by taking the maximum $P_{Correct}$, and the minimum $P_{Correct}$ and linearly normalizing the values to the set weight bound. As a concrete example, to weight a system of values between $0.5-1.5$, the $P_{Correct}$ values over all client nodes are normalized and shifted $+0.5$. In the case of a modest scaling factor between $0.5-1.5$, we observe a convergence at 115 epochs, against the regular 140 epochs, resulting in a $17.858\%$ speedup. In the case of $0.25-1.75$, the speedup is even further improved, converging at the same time as the ideal solution at 80 epochs, a $75\%$ speedup. In evaluating $0.75-1.25$ we notice an improvement in convergence when comparing with $0.5-1.5$ and no weights, with convergence at 105 epochs. With respect to converged error rate, we plot this information within the sub figure of Figure \ref{fig:results_weighted_vqe}. As illustrated, utilizing a weighting system of 0.25-1.75 increases error by $0.33\%$. However, utilizing weights of $0.50-1.50$ allowed for the system to converge $0.49\%$ closer to the ground energy, and $0.37\%$ closer with weights $0.75-1.25$. Increasing range of weights allows for larger steps in parameter updates, which can cause problems of converging on local-minima due to step size being too large. 
Continuing the discussion on machine-bias mitigation, the introduction of weighting further improves the ability of our system to dampen quantum machine noise at real time. Given that we calculate our weight based on noise statistics, any time dependent drift will lower the impact of the quantum machine. Any machine that tends away from being performant will be trusted less and less, until it is calibrated back to a better noise standard. This benefit is very similar to that of boosting in classical machine learning, where better results are amplified and gradually dominate. 

\subsection{Quantum Aproximate Optimization Algorithm} 

\begin{figure}
    \centering
    \includegraphics[width=0.7\linewidth]{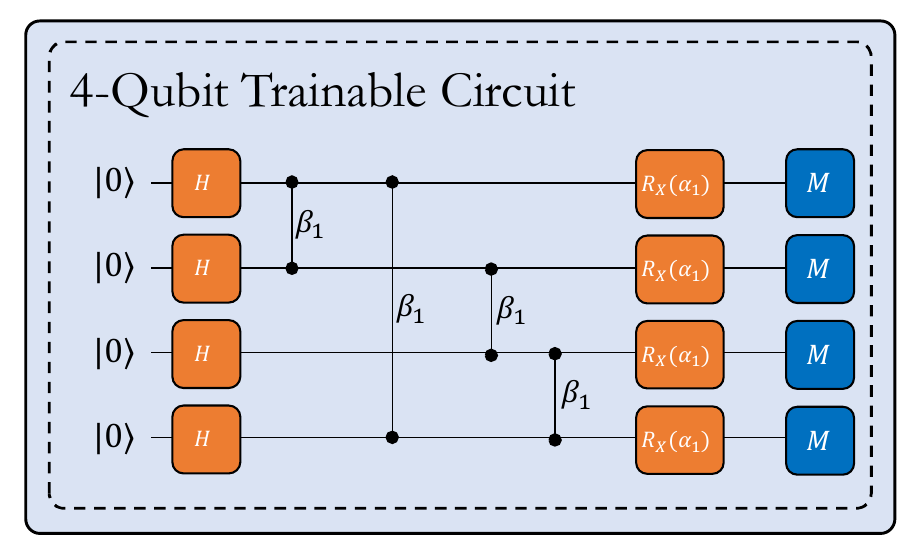}
    \caption{4-Qubit QAOA Circuit. Connected dots represent ZZ gates parameterized by $\beta$. $R_X$ represent rotations around the X-axis by parameter $\alpha$. $H$ represents Hadamard gates. M gates represent measurements.}
    \label{fig:exp_circuit}
\end{figure}

In this section, we demonstrate the Quantum Approximate Optimization Algorithm (QAOA) using a MaxCut problem. QAOA is a quantum algorithm that is used for approximating solutions for combinatorial optimization problems, e.g. MaxCut. In our example, we demonstrate the MaxCut problem over the same graph structure discussed prior with $V=[1,2,3,4]$ and $E=[(1,2),(2,3),(3,4),(1,4)]$. Given a graph G(V,E), where $|V|=n$ and with undirected edge weights of $w_{i,j} > 0 $, the MaxCut problem optimizes towards the partitioning nodes into two sets, such that the number of edges E connecting two nodes $V_i$ and $V_j$ from different subsets is maximized. This is formalized in the characteristic Equation \ref{eqn:maxcut_cost}.
\begin{equation}
    C(x) = \sum_{i,j} w_{i,j}x_i(1-x_j)
    \label{eqn:maxcut_cost}
\end{equation}
Where, in Equation \ref{eqn:maxcut_cost}, $x_i$ is the group of node $i$, and $x_j$ is the group of node $j$. Translating this optimization problem to a diagonal Hamiltonian such that it can be optimized via variational quantum algorithms, it can be mapped as a sum over the edge set $E$.
\begin{equation}
    C(x) = \sum_{(i,j)\subset E}(x_i(1-x_j) + x_j(1-x_i))
    \label{eqn:maxcut_sum}
\end{equation}
Equation \ref{eqn:maxcut_sum} is mapped to a spin Hamiltonian, where $x_i = \frac{1-Z_i}{2}$, as well as swapped from a maximization to a minimization problem, resulting in Equation \ref{eqn:maxcut_ham}
\begin{equation}
    H = -\sum_{j,k)\subset E} \frac{1}{2}(1-Z_jZ_k)
    \label{eqn:maxcut_ham}
\end{equation}
We employ this optimization problem over the graph discussed prior on the IBMQ machines \texttt{Toronto}, \texttt{Santiago}, \texttt{Quito}, \texttt{Lima}, \texttt{Casablanca}, \texttt{Bogota}, \texttt{Manila} and \texttt{Belem} both independently and utilizing the EQC framework. The circuit trained is illustrated in Figure \ref{fig:exp_circuit}, and is motivated by the design outlined in the original QAOA paper \cite{farhi2014quantum}. In this circuit, a total of $2$ parameters are to be optimized, whilst making use of 8 asynchronous systems. In this experiment, we demonstrate the ability of our system to converge with reduced ground state errors at a substantially higher training performance over single machines, highlighting our noise dampening and system throughput contribution.

\begin{figure}
    \centering
    \includegraphics[width=1\linewidth]{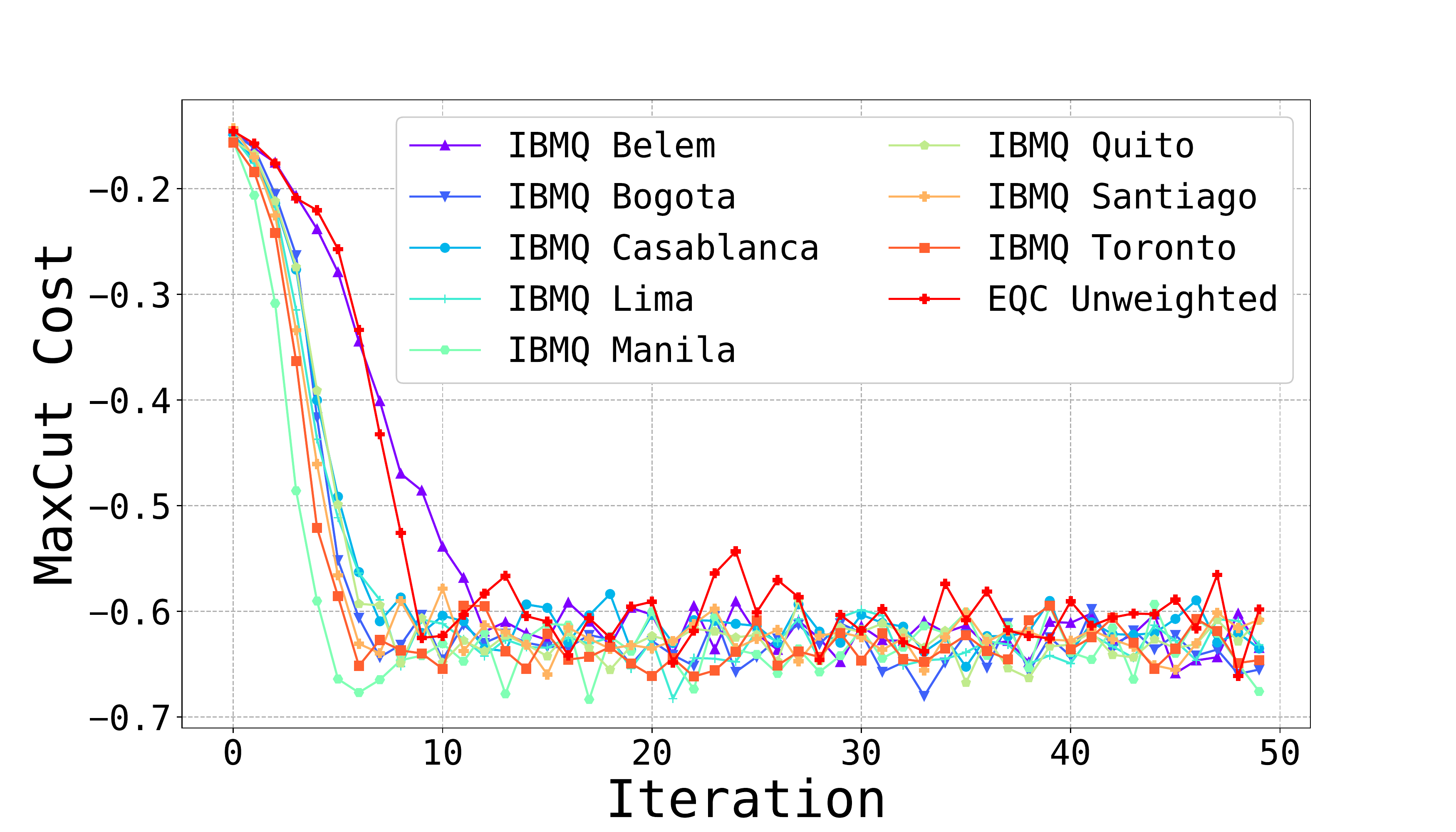}
    \caption{4-Node Unweighted Graph MaxCut Optimization. Shaded regions represent error over 3 experiment runs. Speed represents system data throughput. Graph illustrates comparison between IBMQ independent machines and an unweighted EQC system}
    \label{fig:results_qaoa}
\end{figure}

\begin{figure*}[!t]
    \centering
    \includegraphics[width=1\textwidth]{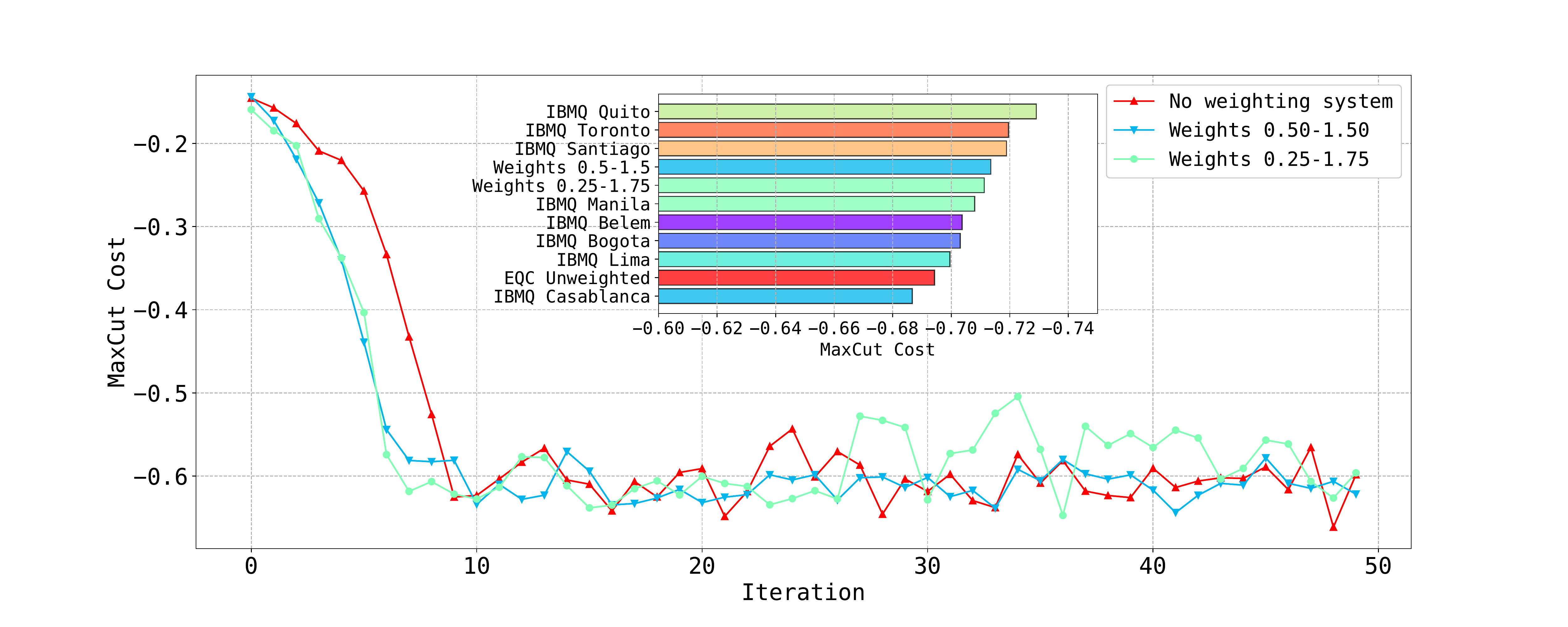}
    \caption{4-Node Unweighted Graph MaxCut Optimization. Shaded regions represent error over 3 experiment runs. Minimum MaxCut Cost represents the lowest MaxCut cost attained. Graph illustrates comparison between unweighted EQC and weighted EQC system. }
    \label{fig:results_weighted}
\end{figure*}

We illustrate the optimization process in Figure \ref{fig:results_qaoa}. EQC converges under similar iterations when compared with the independent quantum processors. Due to experiment run times, there is variation in day to day performance, which can lead to difficulty in directly comparing systems. As a concrete example, Toronto trains over multiple days, with multiple calibration cycles, whereas Belem can can complete training in one hour. EQC performs at a speed of 135,510.2\% faster than the slowest IBMQ machine and 322.4\% faster than the fastest machine. With respect to converged performance with system weighting, we compare the best solutions presented by our models in Figure \ref{fig:results_weighted}. As can be seen, by applying a weighting system based on system quality, the model converges quicker than without, and to a lower final MaxCut cost. Without weighting, EQC attains convergence second to worst, only beating out \texttt{IBMQ Casablanca}. However, by applying our weighting schema, EQC is able to attain solutions close to that of the top 3 IBMQ machines and outperform many worse machines. Weighting approaches improved solutions by 2.863\% for a 0.5-1.5 weighting, and 2.343\% for a 0.25-1.75 weighting. In comparison, using the best machine of \texttt{IBMQ Quito} was a 4.786\% improvement over unweighted EQC. This evidence acts as a strong support for EQC in bolstering distributed quantum computing performance with system-calibration based management. 

\section{Discussion}
EQC has proven to provide a substantial speedup in training variational quantum routines whilst providing a framework for real time noise dampening. By viewing multiple QPUs as an ensemble, we are able to weight QPUs with lower confidence, when they are trapped in a noisy condition, and increase the confidence when they are recovered or calibrated. A good example is the training of VQE on \texttt{IBMQ-Casablanca}. \texttt{IBMQ-Casablanca} was able to train faster than any other QPUs on IBMQ, as shown in Figure~\ref{fig:results}. However, it diverges after being converged for about 20 epochs due to the increasingly deleterious running condition of the machine. For single machine execution, such an impact is inevitable. However, in an ensemble situation, the noise would be recognized and dampened, thereby mitigating this problem. This fluctuating condition of QPU performance highlights the necessity of our ensemble-method for today's erroneous and noisy NISQ devices.  
Additionally, with our weighting system further applied, it is possible to ensemble many noisy quantum machines alongside a few well-behaved QPUs to create a virtualized quantum device with both high-fidelity and high-performance.

\section{Related Work}

Understanding quantum device running behavior and mitigating the errors through algorithms, compilation and architectural approaches is undoubtedly necessary for the success of QC in the NISQ era.

%LaRose \cite{larose2019overview} evaluate four generally used quantum programming environments, including PyQuil, Qiskit, Project and Q\# in terms of library support, hardware, compiler and simulator performance.

Regarding the characterization of NISQ devices, Patel et al. \cite{patel2020experimental} profile the error and execution time of the IBMQ NISQ devices through benchmarkings and draw nine interesting observations. Cross et al. \cite{cross2019validating} define "quantum volume", which quantifies the largest random circuit of equal width and depth that a NISQ device can successfully implement, concerning both gate and SPAM errors. Murali et al \cite{murali2020software} characterize the crosstalk noise of IBMQ devices through random benchmarking, and propose compiler techniques to mitigate the crosstalk impact. Sun and Geller \cite{sun2018efficient} characterize the SPAM error of IBMQ and Rigetti platforms. 

Regarding the error mitigation techniques, Tannu et al. \cite{tannu2019ensemble} observe that the same qubit mapping is often reused for all circuit execution trials, leading to correlated error associated with a certain mapping being amplified to give the incorrect results. Motivated by the concept of diversity, they propose an ensemble of \emph{diverse mappings} (EDM) for different trials of QC execution to avoid the dominance of single error type. Our work is analogous in that we also attempt to alleviate NISQ error through mixture. However, we focus on device-specific bias, and leverage multi-devices simultaneously for VQA training, significantly speedup the execution. Patel and Tiwari \cite{patel2020veritas} focus on estimating correct output from raw erroneous outputs of the NISQ devices through statistical methods. Comparatively, we propose a weighting system to regularize the gradient from the multiple backends based on their device properties and runtime conditions. 

Regarding the acceleration and scaling of NISQ-based QC, Li et al. \cite{li2021software} focus on VQE, and present an application-compiler-hardware codesign for QC program compression with little loss of accuracy. Additionally, they tailor the compiler and architecture according to the problem's Pauli strings, greatly reducing the execution overhead. Patel et al. \cite{patel2020veritas} proposes VERITAS, a distributed quantum computing framework for estimating correct outputs of quantum routines, which can provide substantially useful to single-output-distribution algorithms or tasks on quantum computers.
Tang et al. \cite{tang2021cutqc} introduce \emph{CutQC}, which can decompose a large quantum circuit into several pieces which can then be evaluated on small-scale NISQ devices. The probability distribution of the subcircuits are collected for reconstructing the original distribution. Gokhale et al. \cite{gokhale2019partial} aim at accelerating pulse generation for VQA. They propose to pregenerate the optimal pulses for certain blocks of gate tailored towards VQA circuits, serving as partial compilation for accelerating the classical GRAPE \cite{glaser2015training} based machine control pulse generation. Britt and Humble \cite{britt2017high} investigate the potential architectural design choices, such as CPU/QPU interconnection, for QPU-integrated HPC. Finally, Das et al. \cite{das2019case} propose to multi-program the same NISQ devices for enhancing NISQ device utilization. In order to mitigate the possible interference, three techniques are presented: picking up the reliable regions; unifying the circuit length; monitoring the reliability and disable multi-programming if necessary.
All these techniques, however, are perpendicular to EQC and can be integrated to further mitigate NISQ error, promote utilization rate, and accelerate VQA training speed. For example, if an advanced device (e.g. \texttt{IBMQ Toronto}) can sustain more than one VQA circuit simultaneously, multiple jobs can be distributed to the same backend device for co-execution, further improving the training speed and system utilization.

\section{Conclusions}

The field of adaptive ensembled quantum computing has not been tackled in literature to the best of our knowledge. We present the first framework for adaptive quantum ensemble for more accurate and faster VQA training. Our system is by no means the perfect solution, but serves as a promising start to the idea of ensembled quantum computing (EQC) where various homogeneous or heterogeneous quantum devices can cooperate on a single quantum computing application, harvesting  improved accuracy and superior performance.

%Future developments can look to improve the weighting system, queue management and processor management to maximize the accuracy of EQC, or look at connecting heterogeneous backends from different NISQ vendors for the ensemble.

%\section*{Acknowledgements}
%This material is based upon work supported by the U.S. Department of Energy, Office of Science, National Quantum Information Science Research Centers, Co-design Center for Quantum Advantage ($C^2QA$) under contract number DE-SC0012704. Bo Peng and Karol Kowalski were supported by the "Embedding QC into Many-body Frameworks for Strongly Correlated Molecular and Materials Systems" project, which is funded by the U.S. Department of Energy, Office of Science, Office of Basic Energy Sciences (BES), the Division of Chemical Sciences, Geosciences, and Biosciences. Bo Peng also acknowledged the support by the U.S. Department of Energy, Office of Science, National Quantum Information Science Research Centers. The Pacific Northwest National Laboratory is operated by Battelle for the U.S. Department of Energy under contract DE-AC05-76RL01830.

%%%%%%% -- PAPER CONTENT ENDS -- %%%%%%%%
\section*{Appendix: Convergence Proof}

To show that EQC can converge, we first make the following assumptions: 

\textbf{(I):} The loss function $\ell$ must be a differentiable function with $\ell:\mathbb{R}^d \mapsto \mathbb{R}$, based on the expectation values of observables $\langle O_i\rangle$ from the quantum routine; \textbf{(II):} The application circuit must be in some part comprised of variational gates with some measurements performed, which means
\begin{align*}
\ell([\vec{\theta}]) = \ell(\bra{\psi(\vec{\theta})} O_1 \ket{\psi(\vec{\theta})},\ldots,   \bra{\psi(\vec{\theta})} O_m \ket{\psi(\vec{\theta})} )
\end{align*}
for quantum states $\ket{\psi(\vec{\theta})}$ that are produced using the variational circuit for parameters $\vec{\theta} \in \mathbb{R}^d$ is from a fiducial input state $\ket{0}$; \textbf{(III):} The loss function should be easily computable on a classical computer, and we assume its cost is negligible compared to the time required to estimate $\bra{\psi(\vec{\theta})} O_i \ket{\psi(\vec{\theta})}$.

Under these assumptions, we show that our asynchronous EQC system can converge. This proof is motivated by existing works  \cite{sweke2020stochastic,nedic2001distributed}. If the system is parameterized by $[\vec{\theta}],\theta \subseteq \!R^d$, and some cost function $\ell:\!R^d\to R$, and optimized through stochastic gradient descent rule:
\begin{equation}
    \theta^{t+1}_i = \theta^t_i - \alpha g^t (\theta_i^t)
    \label{eqn:sgd}
\end{equation}
where $g^t$ is an unbiased gradient estimator for the variable $\theta_i$, and hence $\mathbf{E}(g^t(\theta_i^t))=\nabla \ell (\theta_i^t)$. Let's assume $\ell$ is based on a classical observable taking the form of: $\ell([\vec\theta],\langle O_1\rangle_{[\vec{\theta}]},\langle O_2 \rangle_{[\vec{\theta}]}, ... , \langle O_i\rangle_{[\vec{\theta}]})$ where $\langle O_i\rangle_{[\vec\theta]}  $ is the expected value of $O_i$ generated from the circuit parameterized by $[\vec\theta]$. Without losing generality, we assume that for all parameters in $[\vec{\theta}]$, the unbiased estimator of  the partial derivative with respect to $\theta_i$ is calculated using the \emph{parameter shift rule} 
    $\frac{\delta \ell}{\delta \theta_i} = r[\ell(\theta_i+\frac{\pi}{4r} - \ell(\theta_i-\frac{\pi}{4r})]$
where for common operators such as $R_X,R_Y,R_Z$, $r=1/2$. We assume that the system is measuring in a basis comprised of two orthonormal eigenvectors as required by the parameter shift rule, e.g., measuring against $|0\rangle$ and $|1\rangle$.

Since evaluating the exact expectation values from quantum applications becomes rapidly infeasible, we use the sampling mean estimator $\langle o_i \rangle$ of the output $\langle O_i \rangle$. We assume sufficient samples are taken such that $\langle o_i \rangle = \langle O_i \rangle$, and can be used as an unbiased estimator.

Given that we will operate on each parameter $\epsilon$ times, and that the probability that parameter $\theta_i$ fails to be differentiated at any point is $p$ , the probability for a sequence of successes and failures can be expressed as: 
\begin{equation}
    P_{Setting} = (1-p)^{\epsilon-n}p^n
\end{equation}
Based on this, the probability of all possible combinations of failures and successes with a set number of failures is:
\begin{equation}
     P_{n-failures} = \frac{\epsilon!}{(\epsilon-n)!n!}(1-p)^{\epsilon-n}p^n
\end{equation}
The probability of the parameter failing $n$ times or less is:
\begin{equation}
    P=\sum_{j=0}^n\frac{\epsilon!}{(\epsilon-j)!j!}(1-p)^{\epsilon-j}(p^{\epsilon})
\end{equation}
If $p$ is reasonably small, and $\epsilon$ is sufficiently large as in our case, we can see that with $p\to0,\epsilon\to\infty$, $P\to 0$. 

We now discuss the asynchronous SGD. In our case, given the significant difference in QPU execution time, we need to rely on the asynchronous gradient descent update rule:
\begin{equation}
    \theta^{t+1}_i = \theta_i^t - \alpha g^\tau (\theta^\tau_i)
    \label{eqn:asgd}
\end{equation}
where $\tau \leq t$ and $t-\tau$ is equivalent to the time delay of the gradient. We assume the stepsize $\alpha$ is non-increasing, and the time delay $t-\tau$ is bounded by some unknown positive integer $D$ such that the model is only a partially asynchronous method \cite{bertsekas1989parallel}. Consequently, the EQC optimization process follows equal distribution of parameter optimization, as all of the parameters are optimized an equal number of times by cycling over each parameter sequentially, where the system operates on each parameter $1,2,...,m$ and repeats starting from 1 again after operating on $m$. Furthermore, given a sequence obtained by sampling from the cyclic sequence, we denote the sequence used in calculating the updated parameters as $\pi(t)$. Therefore, for some positive integer T, we maintain the validity of $|\pi(t) - t| \leq T$ for $\forall t=1,2,...$ (i.e. the asynchronous nature of the system.)

Let's assume the the gradients are bound by a limit of $C$ which means $||g||\leq C,\ \forall g\subseteq \frac{\delta\ell}{\delta\theta}\forall \theta$. Therefore, the expectations of quantum systems are strictly bound by $0\leq \langle O\rangle \leq 1$ and requiring $\ell(\langle O\rangle)$ to be bound, hence $||g||$ is bounded. We further argue that the difference between updated parameters calculated from time $\tau$ on parameters at time $t$ is bounded by $D$ such that $|t-\tau|\leq D,\  \forall t=0,1,2,...$, etc. Then with respect to \cite{nedic2001distributed}, it can be shown that, if $\ell^* = -\infty$,
\begin{equation}
   \lim_{t\to\infty}\ell([\vec\theta])=-\infty
\end{equation}
otherwise, if $\ell^*$ is finite:
\begin{equation}
    \lim_{t\to\infty}\ell([\vec\theta])= \ell^* + mC^2(\frac{1}{2}+m+2D+T)\alpha
\end{equation}
where $\ell^*$ is the optimal solution to the objective function, which demonstrates the convergence.

\section{Acknowledgements}
This material is based upon work supported by the U.S. Department of Energy, Office of Science, National Quantum Information Science Research Centers, Co-design Center for Quantum Advantage ($C^2QA$) under contract number DE-SC0012704. Bo Peng and Karol Kowalski were supported by the "Embedding QC into Many-body Frameworks for Strongly Correlated Molecular and Materials Systems" project, which is funded by the U.S. Department of Energy, Office of Science, Office of Basic Energy Sciences (BES), the Division of Chemical Sciences, Geosciences, and Biosciences. Bo Peng also acknowledged the support by the U.S. Department of Energy, Office of Science, National Quantum Information Science Research Centers. The Pacific Northwest National Laboratory is operated by Battelle for the U.S. Department of Energy under contract DE-AC05-76RL01830.
%%%%%%%%% -- BIB STYLE AND FILE -- %%%%%%%%
\bibliographystyle{IEEEtranS}
\bibliography{refs}
%%%%%%%%%%%%%%%%%%%%%%%%%%%%%%%%%%%%

\end{document}